\documentclass{JHEP3}
\usepackage{graphicx}
\usepackage{epsfig,amsmath,amsthm}
\newcommand{\be}{\begin{equation}}
\newcommand{\ee}{\end{equation}}
\newcommand{\bea}{\begin{eqnarray}}
\newcommand{\eea}{\end{eqnarray}}
\def\bse{\begin{subequations}}
\def\ese{\end{subequations}}

\def\IZ{\relax\ifmmode\hbox{Z\kern-.4em Z}\else{Z\kern-.4em Z}\fi}

\newcommand{\non}{\nonumber \\}

\def\bi{\begin{itemize}} \def\ei{\end{itemize}}

\def\Schw{Schwarzschild }
\def\({\left(} \def\){\right)}
\def\[{\left[} \def\]{\right]}
\def\risco{r_{\rm ISCO}}

\title{ \center{Post-ISCO Ringdown Amplitudes in Extreme Mass Ratio Inspiral}}
\author{Shahar Hadar and Barak Kol\\
Racah Institute of Physics, Hebrew University\\
Jerusalem 91904, Israel\\
E-mail: {\email{shaharhadar@phys.huji.ac.il},
\tt\href{mailto:barak_kol@phys.huji.ac.il}{barak\_kol@phys.huji.ac.il}}
}

\abstract{An extreme mass ratio inspiral consists of two parts:
adiabatic inspiral and plunge. The plunge trajectory from the
innermost stable circular orbit (ISCO) is special (somewhat
independent of initial conditions). We write an expression for its
solution in closed-form and for the emitted waveform. In
particular we extract an expression for the associated black-hole
ringdown amplitudes, and evaluate them numerically.}
\begin{document}

\section{Introduction and summary}

Gravitational wave observatories (see for example the reviews
\cite{IFO-rev} and references therein) demand knowledge of the
waveform emitted by an inspiralling binary system of compact
objects. Analytic control is possible in two limits. The first is
the \emph{Post-Newtonian (PN) approximation}, which holds whenever
the velocities are small compared with the speed of light. The
second limit, the one relevant to this paper, is that of an
\emph{extreme mass ratio (EMR)}, for example, a compact projectile such
as a stellar size black hole (BH) falling into a super-massive
black hole.
 Numerical solutions complement the analytic methods and have become
possible in recent years - see the review \cite{PretoriusRev} and
references therein. However, while numerical simulations work very
well for comparable masses, they encounter problems in the EMR case due to the existence of two disparate scales
which proves problematic for the discretized grid. So it happens that just where the numerical method loses precision, the analytic EMR approximation becomes more effective. Moreover, analytic methods improve insight into the problem, especially into its dependence on parameters.

The evolution of an Extreme Mass Ratio Ispiral (EMRI) is
customarily divided into two stages: an \emph{adiabatic inspiral}
where the system moves on quasi-bound orbits and slowly loses
energy to gravitational waves, and a \emph{plunge} phase where the
system is set on a course of collision even when the self-force is
neglected. In this limit the plunge phase simplifies to consist of
geodesic motion of the projectile till it reaches the horizon of
the larger BH. The ISCO (innermost stable circular orbit) is the
border between the two stages, after which the compact object
freely falls into the horizon, followed by merger and ringdown
(through quasi-normal modes) into a new stationary state. The
smooth transition between adiabatic inspiral and plunge was
beautifully understood in \cite{Ori:2000zn}. While the adiabatic
inspiral is well within the applicability domain of the heavily
studied PN approximation, known today up to order 3.5PN (as
summarized in the review \cite{BlanchetRev}) considerably less is
known analytically about the plunge phase which is the subject of
this paper.

We would like to have a mathematical description of the waveform
throughout its evolution, beginning with the adiabatic inspiral
and lasting until the waveform vanishes after the black hole rings
down and reaches steady state.

The current state-of-the-art for such an all-time analytic expression
goes under the name ``Effective One Body formalism'' (EOB)
\cite{EOB1,EOB2}, see also the recent review
\cite{EOB-rev}.\footnote{The method's main idea is an attempt to
carry over to General Relativity (GR) the successful classical
reduction of the two body problem to a reduced one body problem,
where here in GR the reduced point particle moves in some
``corrected'' metric background. This background does not solve
Einstein's equations and as such seems to offer enough adjustable
parameters to fit for the sought waveforms. Its proven success
makes it undisputable as an ad-hoc method. Yet it is not obvious
that it is theoretically justified or natural, in the sense that a
natural approximation would not contain unnecessary adjustable
parameters nor would it lack parameters for higher order
approximations.} The EOB approach combines quite a number of ideas
and ingredients, but here it will suffice to concentrate on its
handling of the ringdown phase, which is quite independent of the
other ingredients, and is based on the following assumption: the
modeled waveforms during the adiabatic and the ringdown phases may
be matched together without the need for an independent modeling
of an intermediate phase. This assumption was later vindicated by
full numerical simulations. More specifically, the adiabatic phase
is carefully modeled
% \cite{DamourNagarEMRI} p.3 dissipative force \hat{\cal F}_\phi as a function of v_phi is resumed to have a pole at ISCO.
and a general expression for the ringdown phase is matched
directly onto it. In \cite{NDT-EMRI} the full waveform was
computed in the extreme mass ratio limit by numerically evolving
the trajectory within the EOB approach and numerically determining
the radiation.  In \cite{DamourNagarEMRI}
the model evolved to its present form where the waveform emitted during ringdown is
modeled by the most general ringdown signal -- a linear superposition of its quasi-normal modes (eq. (18) there) \be % \cite{DamourNagarEMRI})
 \Psi^{\rm ringdown}_{22}(t)= \sum_n\, R^+_n\, e^{-\sigma^+_n\, t} + \sum_n\, R^-_n\, e^{-\sigma^-_n\, t}~, \label{general-ringdown}
\ee
 where $\sigma_n^\pm \equiv \alpha_n \pm i \omega_n$ are positive/negative quasi-normal-mode (QNM) complex
frequencies, $R^\pm_n$ are the corresponding ringdown amplitudes, and $n$ labels the QNM overtone number. This matching is implemented by matching the EOB model for the adiabatic phase with a general ringdown waveform (\ref{general-ringdown}) \footnote{With $R^-_n=0$.} around some judiciously chosen matching time,\footnote{The crossing of the ``light ring'' at $r=1.5 r_s$.}
 matching at as many time points as the number of amplitudes to be determined.

In this paper we shall compute the ringdown amplitudes directly
from the theory for a certain special plunge trajectory.\footnote{ A similar approach was taken in \cite{MinoBrink}. However they differ significantly from the current study both in focus and in results. In particular, they do not calculate the ringdown amplitudes.}
For simplicity we consider the trajectory to be moving in the
background of a non-rotating (Schwarzschild) BH, and we moreover
consider the plunge trajectory to be the one which starts at the
innermost stable circular orbit (ISCO) at $\risco=3 r_s$ (in \Schw
coordinates, where $r_s$ is the \Schw radius). This trajectory is
special since the eccentricity of an orbit is known to decrease during the Keplerian
regime $r>>r_s$ of the inspiral \cite{Peters}, thus being a ``Keplerian attractor'' and hence only weakly dependent on initial conditions. In the case that the inspiral has been going on for enough time for
the eccentricity to be essentially radiated away by the time ISCO
is reached, then the plunge trajectory will indeed start from roughly ISCO. However,
there are relatively recent indications that in some probable astrophysical scenarios other initial conditions are expected, namely initial conditions such that the last stable orbit is not circular, see for example \cite{BarackCutler,HopmanAlexander,AmaroSeoane:2007aw}. Yet, in this paper we confine ourselves to the post-ISCO plunge for its simplicity and for being a Keplerian attractor.

For the post-ISCO plunge trajectory we \emph{compute the
form of the emitted gravitational waves}. In the $t \to
\infty$ limit we may \emph{extract the post-ISCO ringdown
amplitudes}, while the formal limit $t \to -\infty$ corresponds to
radiation from ISCO.
% PRDv2
We wish to stress that the waveform under study continuously
interpolates between two time regions: in early time the compact
object is near ISCO and the radiation is characterized by the
frequency of that trajectory, in the transition period the
trajectory is not periodic and hence there are no sharp
frequencies in the waveform, while at late times the decaying
radiation is characterized by the (complex) frequencies of the
quasi-normal modes which are set by the black hole, and not by the
trajectory.

In section \ref{sec:orbit} we solve analytically for the plunge
trajectory, namely the trajectory with energy and angular momentum
identical to the ISCO values. The expression for $r=r(\phi)$ is
known (at least since \cite{Chandrasekhar:1985kt}) to be
especially simple and the time dependence can also be expressed in
closed form (\ref{t-and-r},\ref{t-and-phi}).
In section \ref{sec:source} we review the Zerilli/Regge-Wheeler
\cite{Regge:1957td,Zerilli:1970se} theory for radiation in the
background of a \Schw BH for a given source of energy-momentum
(using Martel-Poisson \cite{Martel:2005ir}). We proceed to
substitute in the energy-momentum for our post-ISCO plunge trajectory.
In section \ref{sec:quasi} we write a formal solution to the wave
equation in terms of Green's functions (\ref{time domain solution for master function}). We extract the late-time
ringdown form by deforming a frequency integral into the complex
domain and transforming it to a sum over residues at the
quasi-normal modes. The final expression for each post-ISCO
ringdown amplitude (\ref{amplitude def}) calls for solving certain ordinary differential
equations (the radial wave equations) at the QNM frequencies and
performing a certain weighted radial integral over the source.
In section \ref{sec:numerics} we numerically evaluate these integrals
 and obtain the amplitudes of the leading ringdown modes shown in tables \ref{table of amplitudes 1},\ref{table of amplitudes 2} and in figure \ref{fg:amplitude listplot}. These values were confirmed by full numerical simulations \cite{InProgress}.

Our expression (\ref{time domain solution for master function}) contains more than the ringdown amplitudes: it describes \emph{the
full plunge waveform as a function of time}.
Solving for all $(l,m)$ is equivalent to solving an inhomogeneous 3+1 wave equation
and extracting the asymptotic outgoing signal.

The impact of our work is that hereafter the EMRI post-ISCO
ringdown amplitudes should be considered to be determined and
known (at least under the assumptions above) and any waveform
model must conform with it at $t \to \infty$ (and even as early as
ISCO is crossed assuming the full plunge waveform is used rather
than only its late time part).

\emph{Generalizations}. Several generalizations remain. First, one
may consider the background to be a rotating BH (Kerr), in which
case there are two kinds of post-ISCO plunge trajectories:
co-rotating and counter-rotating, and the Zerilli/Regge-Wheeler
equations must be replaced by the Teukolsky equation. Second, one may consider orbits where eccentricity (and being off
the equator plane in the Kerr background) was not washed away yet.
Finally higher order effects may be incorporated such as
accounting for the projectile's spin.

%----------------------------------------------------------------------%
%----------------------------------------------------------------------%
\section{Plunge trajectory}
\label{sec:orbit}
%----------------------------------------------------------------------%
%----------------------------------------------------------------------%

Consider an inspiralling binary system in the Extreme Mass Ratio
Inspiral (EMRI) limit, namely $\tilde{m} \ll M$ where $\tilde{m}$
is the mass of the compact object falling into a BH with mass M,
or Schwarzschild radius $r_{s}=2 G M$.\footnote{We reserve the
letter $m$ to denote the magnetic number - an index for the
spherical harmonic functions $Y^{lm}$.} In order to calculate the
gravitational radiation emitted from a system we need to find the
dynamics of the compact source. For reasons to be discussed below
we consider the (geodesic) trajectory which spirals out of the
innermost stable circular orbit (ISCO) at $\risco=3 r_s$ and
plunges into the black hole. In this section we shall solve for
this trajectory.

We concentrate on the case when the inspiral has been going on long enough for the
radiation reaction force to have circularized the trajectory by
the time the compact object reaches the innermost stable circular
orbit (ISCO) at $\risco=3 r_s$. This assumption deserves a
discussion. Peters \cite{Peters} found that in the Keplerian
regime $r \gg \risco$ energy and angular momentum are lost such
that eccentricity decreases with the following rate \be
 p=c_0\, e^\frac{12}{19} \, \( 1 + \frac{121}{304}
 e^2\)^{\frac{870}{2299}} \label{e-rad}\ee
 where $p$ is the Keplerian semi-latus rectum, and $e$ is the
eccentricity, such that the trajectory is given by
$r(\theta)=p/\(1+e \cos(\theta) \)$ and the semi-major axis $a$ is
given by $a=p/(1-e^2)$. Assuming small $e$ and extrapolating
(\ref{e-rad}) down to ISCO we can estimate that \be
 e_f \simeq e_i \( \frac{\risco}{p_i} \)^{\frac{19}{12}} ~,\label{e-rad2} \ee
where $f,i$ subscripts denote final and initial, respectively.
Note that this is only an approximation since the radiation
reaction force is different in the strong gravity regime $r
\gtrsim \risco$, and it is actually known that $e$ increases
somewhat just before ISCO \cite{AKPO}. Altogether, (\ref{e-rad})
and (\ref{e-rad2}) quantify the rate of circularization and
explain why $e$ would be small at ISCO given $p_i \gg \risco$.
Hence ISCO is a Keplerian attractor. Yet, as mentioned in the introduction it should be borne in mind
that initial conditions such that the last stable orbit is not circular are also to be expected.

Since ISCO is a (marginally) unstable orbit, we choose the
plunging orbit with the same conserved parameters as those at
ISCO, namely \bea
 \tilde{E} &=& \tilde{E}_\mathrm{ISCO}\equiv \frac{2 \, \sqrt{2}}{3} \non
 \tilde{L} &=& \tilde{L}_\mathrm{ISCO} \equiv \sqrt{3} \, r_{s} ~,\eea
 where $\tilde{E}_\mathrm{ISCO}$, $\tilde{L}_\mathrm{ISCO}$ are the
particle's energy per unit mass and angular momentum per unit mass
at ISCO. During the plunge radiation reaction can be neglected as
the object is already falling ``without its help''.\footnote{The
leading corrections for the ISCO values at plunge were computed in
\cite{Ori:2000zn} eq. (3.26) and they were found to be
proportional to $\( \tilde{m}/M \)^{4/5}$.}

We wish to find explicitly the trajectory of this free-fall.
Without loss of generality the trajectory can be considered to lie
in the equatorial plane and the equations governing the object's
motion are
\bea
 \tilde{E} &=& f(r)\, \frac{dt}{d\tau}
\label{geodesic motion equations-tdot}\\
 \tilde{L} &=& r^{2}\, \frac{d\phi}{d\tau}
\label{geodesic motion equations-phidot}\\
 \tilde{E}^{2} &=& \left(\frac{dr}{d\tau}\right)^{2}+f(r)(\tilde{L}^{2}/r^{2}+1)
\label{geodesic motion equations-rdot} ~,\eea
where $f(r)=1-\frac{r_{s}}{r}$.\\
In our special orbit, there is a simple relation between the $r$
and $\phi$ coordinates of the particle (noticed at least as early
as \cite{Chandrasekhar:1985kt}) \be
 \frac{\risco}{r} = 1 +\frac{12}{(\phi-\phi_0)^{2}}
  \label{r and phi} ~,\ee
 where $\phi_0$ is an arbitrary constant
(sometimes we will choose it to be zero). We will also use
conventions such that $\frac{d\phi}{d\tau}>0$.

We proceed to solve for the relation between $t$ and $r$ through
integration of equation (\ref{geodesic motion equations-rdot}),
using equation (\ref{geodesic motion equations-tdot}) to change
the derivative with respect to $\tau$ to a derivative with respect to $t$. For a generic orbit we can
solve for $t=t(r)$ by quadrature involving an elliptic integral
\be
 \tilde{E}\int{\frac{dr}{f(r) \, \sqrt{\tilde{E}^{2}-f(r)\left( \, \frac{\tilde{L}^{2}}{r^{2}}+1 \, \right)}}}=-\int{dt}
 \label{elliptic integral} ~.\ee
For our special orbit, the expressions simplify due to the special
form of the effective potential (namely that both its first \emph{and}
second derivatives vanish at $r_\mathrm{ISCO}$). Plugging in our
constants of motion, $\tilde{E}_\mathrm{ISCO}$ and
$\tilde{L}_\mathrm{ISCO}$, it can be seen that (\ref{geodesic
motion equations-rdot}) becomes
 \be
 \( \frac{dr}{d\tau} \)^2 = \frac{1}{3^2}\, \(
 \frac{\risco}{r}-1 \)^3
 \label{simplification of elliptic integral} ~.\ee

%----------------------------------------------------------------------%
\begin{center}
\begin{figure*}[b!]
%\setlength{\tabcolsep}{ 40 pt }
%\footnotesize{
        \begin{tabular}{cc}
            (a) & (b) \\
            \includegraphics[width=7cm,height=7cm]{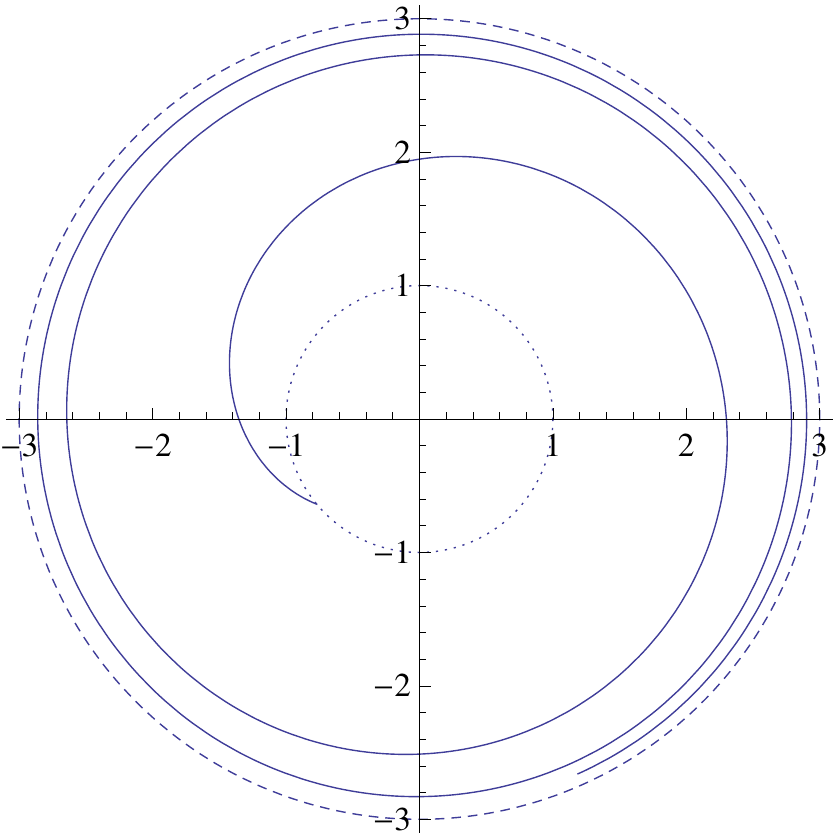} &
            \includegraphics[width=7cm,height=7cm]{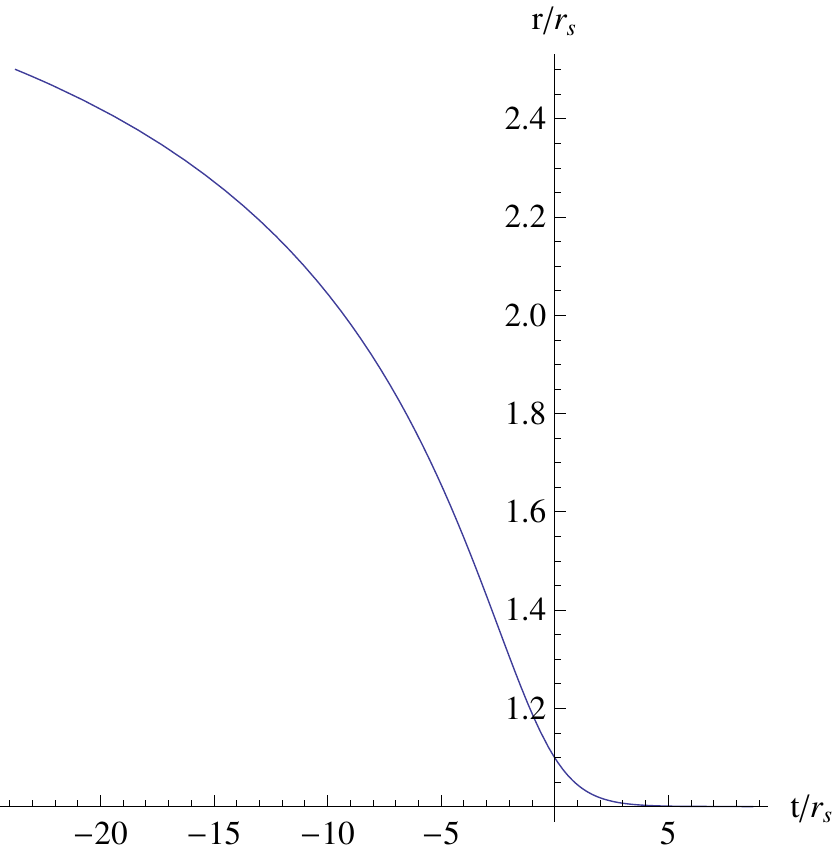}\\
        \end{tabular}
%}
 \caption{(a) Spatial trajectory of the plunging particle,
$r(\phi)$, normalized so that $r_{s}=1$. The dashed line at $r=3$
represents the ISCO, whereas the dotted line at $r=1$ represents
the horizon. (b) $r$ coordinate of the plunging particle as a
function of coordinate time $t$ (equation (\protect\ref{t-and-r})).}
\label{fg:plunge trajectory}
\end{figure*}
\end{center}
%%----------------------------------------------------------------------%

This obviously simplifies the integrand, and with a change of
variables \\ $\chi:=\frac{1}{2}\left(\frac{\risco}{r}-1\right)$ it is
brought to the form of a rational function. Evaluating the
integral, we find that our special orbit satisfies the following
relations \bea
 t(r)/r_s &=& \frac{2 \, r\left( \, 1-12 \, \frac{r_{s}}{r} \, \right)}{r_{s}\sqrt{\chi}}-22\sqrt{2} \, \tan^{-1}\left(\sqrt{2\chi}\right)+
 2 \, \tanh^{-1}\left(\sqrt{\chi} \right) + t_{0}
\label{t-and-r} \\
 t(\phi)/r_s &=& 3\sqrt{6} \, (\phi-\phi_0)\left(1+\frac{4}{(\phi-\phi_0)^{2}+12}\right)-22\sqrt{2} \, \tan^{-1}\left(\frac{2\sqrt{3}}   {\phi_0-\phi}\right)\\
&+& 2 \, \tanh^{-1}\left(\frac{\sqrt{6}}{\phi_0-\phi} \right) + t_{0} \nonumber\\
\label{t-and-phi} \eea where $t_{0}$ is an arbitrary constant
which we shall fix in section \ref{sec:numerics}. We note that
$\chi$ is closely related to the standard change of variables in
Kepler's problem $u:=1/r$ and that $\chi$ varies from $\chi=0$ at
ISCO to $\chi=1$ at the horizon.

%----------------------------------------------------------------------%
%----------------------------------------------------------------------%
\section{Source for gravitational waves}
\label{sec:source}
%----------------------------------------------------------------------%
%----------------------------------------------------------------------%

In this section we proceed to describe the equation satisfied by
the emitted gravitational waves given a source moving on the
geodesic trajectory described in the previous section.

\subsection{A general source in the Schwarzschild background}
\label{Schwarzschild Black Hole Perturbations With a General
Source}

The theory of linearized perturbations of Schwarzschild space-time
was initially developed in the pioneering works of Regge and
Wheeler (1957) \cite{Regge:1957td} and of Zerilli (1970)
\cite{Zerilli:1970se}. Essentially, when expanding the metric
perturbations in spherical harmonics, the physics of the
perturbations reduces to the physics of a $(1+1)$ dimensional wave
equation with a nontrivial potential, satisfied by two ``master
functions'' (for odd and even parities) constructed from the
metric. These master functions characterize the metric
perturbations.

A convenient formalism, suitable for the problem of analyzing
gravitational waves generated by material sources, was introduced
by Martel and Poisson (2005) \cite{Martel:2005ir}. That paper
presents gauge-invariant and covariant form of the Regge-Wheeler
equation (describing odd-parity perturbations of Schwarzschild
space-time) and of the Zerilli equation (describing the even
parity perturbations) with two corresponding scalar master
functions and invariant source terms constructed from the
stress-energy tensor of the matter responsible for the
perturbation. In this subsection we review the general theory of
Schwarzschild BH perturbations with sources, closely following
\cite{Martel:2005ir} (see also \cite{MSSTT}).

The Schwarzschild metric can be written as
 \bea
 ds^2 = g_{\mu\nu}\, dx^\mu dx^\nu &=& g_{ab}\, dx^a dx^b + r^2 \, \Omega_{AB}\,
 d\theta^A d\theta^B
 \label{schw metric.1}\\
 g_{ab}\, dx^a dx^b&=&-f \, dt^2+f^{-1} \, dr^2
 \label{schw metric.2}\\
 \Omega_{AB} \, d\theta^A d\theta^B&=&d\theta^2+\sin^2\theta \, d\phi^2
 \label{schw metric.3}
 ~,\eea
where upper-case latin indices run over the values 2,3
($\theta,\phi$) and lower-case latin indices run over the values
0,1 ($t,r$). We study small perturbations around the Schwarzschild
space-time, so it is useful to write the metric in the form
$g_{\mu\nu} = g_{\mu\nu} + h_{\mu\nu} \label{metric+pert},
\left|h_{\mu\nu}\right| \ll 1$. Next, we expand the perturbation
in spherical harmonics, according to the transformation laws of
the different components under rotation. For the odd-parity
sector \bea
 h_{ab} &=& 0 \label{odd expansion 1} \\
 h_{aB} &=& \sum_{lm} h^{lm}_a \, X_B^{lm}
 \label{odd expansion 2} \\
 h_{AB} &=& \sum_{lm} h^{lm}_2 \, X_{AB}^{lm} \label{odd expansion 3}
~,\eea
 where $X_B^{lm}$ and $X_{AB}^{lm}$ are the odd-parity vector
and tensor spherical harmonics\footnote{ More details regarding
the spherical harmonics can be found in \cite{Martel:2005ir}, and
in particular we record their normalization conventions:  $\int
\bar{Y}_{lm} Y_{l'm'}\, d\Omega = \delta_{ll'}\delta_{mm'}$, $\int
\bar{Y}^A_{lm} Y^{l'm'}_A\, d\Omega = l(l+1)\,
\delta_{ll'}\delta_{mm'}$, $\int \bar{X}^A_{lm} X^{l'm'}_A\,
d\Omega = l(l+1)\, \delta_{ll'}\delta_{mm'}$, $\int
\bar{Y}^{AB}_{lm} Y^{l'm'}_{AB}\, d\Omega = \frac{1}{2}
(l-1)l(l+1)(l+2)\, \delta_{ll'}\delta_{mm'}$ and finally $\int
\bar{X}^{AB}_{lm} X^{l'm'}_{AB}\, d\Omega = \frac{1}{2}
(l-1)l(l+1)(l+2)\, \delta_{ll'}\delta_{mm'}$.},
 respectively. For
the even-parity sector \bea
 h_{ab} &=& \sum_{lm} h^{lm}_{ab} \, Y^{lm}
\label{even expansion 1} \\
 h_{aB} &=& \sum_{lm} j^{lm}_a \, Y_B^{lm} \label{even expansion 2} \\
 h_{AB} &=& r^2 \sum_{lm} \bigl( K^{lm} \, \Omega_{AB} \, Y^{lm} + G^{lm}
\, Y_{AB}^{lm} \bigr) \label{even expansion 3} ~,\eea
 where $Y^{lm}$,
$Y_B^{lm}$ and $Y_{AB}^{lm}$ are the even-parity scalar, vector
and tensor spherical harmonics, respectively. Plugging these
expressions into the Einstein equations and using the
orthogonality relations of the spherical harmonics, a wave
equation for every $(l,m)$ is obtained, satisfied by the
corresponding master functions. The Cunningham-Price-Moncrief
function, characterizing odd-parity perturbations, is defined by
\be
 \psi^{lm}_{\rm o} := \frac{2r}{(l-1)(l+2)} \, \varepsilon^{ab}
\Bigl( \partial_a \, h^{lm}_b - \frac{2}{r} \, r_a \, h^{lm}_b
\Bigr)
 \label{cpm function} ~,\ee
  where $\varepsilon^{ab}$ is the
Levi-Civita tensor on the $(t,r)$ manifold and $r_a=(0,1)$ is the
unit co-vector in the $r$ direction. The Cunningham-Price-Moncrief
function satisfies the Regge-Wheeler equation \be
 ( \Box - V_{\rm o}^{l} ) \, \psi_{\rm o}^{lm} = S_{\rm o}^{lm} \label{rw equation} ~,\ee
  where \be
  V_{\rm o}^{l} = \frac{l(l+1)}{r^2} - \frac{3r_{s}}{r^3} \label{rw potential}
 \ee
is the Regge-Wheeler potential, and $S_{\rm o}^{lm}$ is the source
term, on which we will focus later.

The Zerilli-Moncrief function, characterizing even-parity
perturbations, is defined by \be
 \psi^{lm}_{\rm e} :=
\frac{2r}{l(l+1)} \biggl[ \tilde{K}^{lm} + \frac{2}{\Lambda}
\bigl( r^a r^b \left(h^{lm}_{ab}-2 \, \nabla_a\left(j_b - \frac{1}{2}
\, r^2 \, \nabla_b G\right)\right) - r \, r^a \nabla_a \tilde{K}^{lm}
\bigr) \biggr] \label{zm function}
 ~,\ee where \be
  \Lambda := (l-1)(l+2) + \frac{3r_{s}}{r} \label{lambda def}
 \ee
 \be
 \tilde{K} := K + \frac{1}{2} \, l(l+1) \, G - \frac{2}{r} \, r^a \left(j_a
- \frac{1}{2} \, r^2 \, \nabla_a G\right) \label{k def}
 ~,\ee and
$r_a=(0,1)$ is the unit co-vector in the $r$ direction on the
$(t,r)$ submanifold. The Zerilli-Moncrief function satisfies the
Zerilli equation \be
 ( \Box - V_{\rm e}^{l} ) \, \psi_{\rm e}^{lm} =
S_{\rm e}^{lm} \label{zer equation}
~,\ee where \be
 V_{\rm e} = \frac{1}{\Lambda^2} \, \biggl[ \mu^2 \biggl(
  \frac{\mu+2}{r^2} + \frac{3r_{s}}{r^3} \biggr)
+ \frac{9 \, r_{s}^2}{r^4} \biggl(\mu + \frac{r_{s}}{r} \biggr)
\biggr]
 \label{zer potential} \ee
  is the Zerilli potential ($\mu:=(l-1)(l+2)$) and
$S_{\rm e}^{lm}$ is the source term.

The source terms, as was earlier stated, are constructed from the
stress-energy tensor of the material source of gravitational
waves. In the odd-parity sector, the source term is given by \be
 S_{\rm o}^{lm}=- \frac{2r}{(l-1)(l+2)} \, \varepsilon^{a b} \, \nabla_a
P_b
 \label{rw general source term} ~,\ee where
 \be
  P^a = \frac{16\pi \, r^2}{l(l+1)} \int T^{aB} \, X^{* lm}_B \, d\Omega
\label{odd source p} ~,\ee
 and $X^*$ denotes the complex conjugate of $X$.

In the even-parity sector, the source term is given by \bea
 &S_{\rm e}& = \frac{4}{\Lambda} \, r_a Q^a - \frac{1}{r} \, Q^\sharp +
\frac{2}{(\mu+2) \, \Lambda} \biggl\{ - 2 \, r^2 \, r^a \nabla_a Q +
\frac{12r_{s}}{\Lambda} \, r_a r_b Q^{ab} + 2 \, r \, f \, Q^\flat \nonumber\\
 &+& \frac{r}{\Lambda} \biggl[ \mu \, (\mu-2) + 6 \, (\mu-3) \,
\frac{r_{s}}{r} + 21 \, \frac{r_{s}^2}{r^2} \biggr] Q \biggr\}
 \label{general even source term} ~,\eea
  where \bea Q^{ab} &=& 8\pi
\int T^{ab} \, Y^{*lm} \, d\Omega
\label{q even source term.1} \\
Q^a &=& \frac{16\pi \, r^2}{l(l+1)} \int T^{aB} \, Y^{*lm}_B \, d\Omega
\label{q even source term.2} \\
Q^\flat &=& 8\pi \, r^2 \int T^{AB} \, \Omega_{AB} \, Y^{*lm} \, d\Omega
\label{q even source term.3} \\
Q^\sharp &=& \frac{32\pi \, r^4}{(l-1)l(l+1)(l+2)} \int T^{AB} \,
Y^{*lm}_{AB} \, d\Omega \qquad \label{q even source term.4} \\
Q=g^{ab}Q_{ab} \label{q even source term.5}
~.\eea

In this work we ran a check, independent of the derivation of
results in \cite{Martel:2005ir}, of the expressions for the
source terms (\ref{rw general source term}) and (\ref{general even
source term}). The idea is simple - we checked that both sides of
the Regge-Wheeler and Zerilli equations with source terms,
equations (\ref{rw equation}) and (\ref{zer equation}), agree -
when written explicitly as functions of the coefficients of the
spherical harmonics in the decomposition of the metric
perturbations (e.g. $h^{lm}_a$,...) appearing in equations
(\ref{odd expansion 1}) - (\ref{even expansion 3}). When
substituting expressions (\ref{cpm function}) and (\ref{zm
function}) in the left hand sides of the equations (respectively)
such an explicit expression is obtained. For the right hand sides
of (\ref{rw equation}), (\ref{zer equation}), we used relations
between the coefficients in the spherical harmonic decomposition
of the stress-energy tensor ($P^a$ in (\ref{odd source p}), the
$Q$'s in (\ref{q even source term.1})-(\ref{q even source
term.4})) and the metric perturbation coefficients ($h^{lm}_a$,
etc..). These relations can be obtained by substituting the
decompositions to spherical harmonics in the Einstein equation and
equating the different spherical harmonic coefficients. Although
the idea is simple, its execution demands some tedious algebra.
The process described above was carried out for equations (\ref{rw
equation}), (\ref{zer equation}), and the two sides of both
equations were found to coincide.

Combining the two wave equations (\ref{rw equation}), (\ref{zer
equation}) into a single notation and making the wave operator
explicit, we obtain in the frequency domain \be
 \left[\frac{\partial^2}{\partial r_{*}^{2}}+\omega^{2}-f \, V_{\rm
o/e}^{l}(r)\right]\psi_{\rm o/e}^{lm}(r,\omega)=f \, S_{\rm
o/e}^{lm}(r,\omega) \label{eqn:wave} ~,\ee
 where
$r_{*}=r+r_{s}\log(r/r_{s}-1)$ is the usual tortoise coordinate
and $\psi_{\rm o/e}^{lm}(r,\omega)$ and $S_{\rm
o/e}^{lm}(r,\omega)$ are the Fourier transforms of the master
function and source term, respectively. Notice that changing to
derivative with respect to $r_{*}$ instead of $r$ casts the
equation explicitly in the form of a flat space $1+1$ dimensional
wave equation with a potential.
 The integration of these equations will be discussed in section
\ref{sec:quasi}.

For later use we record the relation between the master functions
and the transverse-traceless part of the metric perturbation
(which is gauge-invariant) at future null infinity \bea
 h_{ab} &=& 0
\label{metric perturbation at null infinity - ab} \\
 h_{aB} &=& 0
\label{metric perturbation at null infinity - aB} \\
 h_{AB} &=& r\sum_{lm} \left( \psi_{\rm e}^{lm} Y^{lm}_{AB} +
\psi_{\rm o}^{lm} X^{lm}_{AB} \right) \label{metric perturbation
at null infinity - AB} ~.\eea

\subsection{Plunging mass as a source}
\label{Plunging Mass as Source of Gravitational Waves}

We turn to calculate the source terms corresponding to our
plunging point particle.  The stress-energy tensor corresponding
to a point-like mass is given by \be
 T^{\mu\nu} = \tilde{m} \int d\tau\ (-g)^{-1/2}\
u^{\mu}u^{\nu}\
\delta^{(4)}\left(x^{\alpha}-x^{\alpha}_{p}(\tau)\right)
\label{general stress-energy tensor} ~,\ee where
$\delta^{(4)}(x^{\alpha}-y^{\alpha})$ is the four-dimensional
Dirac functional, $u^{\mu}$ is the 4-velocity and
$x^{\alpha}_{p}(\tau)$ is the location of the point particle which
approximates the compact object. $x^{\alpha}_{p}(\tau)$ are the
same functions discussed in section \ref{sec:orbit} and the added
subscript $p$ is to distinguish them in the current context.

In the Schwarzschild space-time, after localizing the $\tau$
integral onto $t_{p}(\tau)$ \be
T^{\mu\nu}=\tilde{m} \, \frac{f}{\tilde{E}} \, \frac{u^{\mu}(t)u^{\nu}(t)}{r^{2}} \, \delta\left(r-r_{p}(t)\right)
 \, \delta\left(\theta-\frac{\pi}{2}\right) \, \delta\left(\phi-\phi_{p}(t)\right)
\label{schw stress-energy tensor} ~.\ee
 To compute the odd-parity
source term, we first need to calculate $P^a$ (\ref{odd source
p}). The integral over the angles is trivial due to the delta
functions and we obtain \be
P^a=\frac{16\pi \, \tilde{m} \, \tilde{L} \, f(r_p)}{l(l+1) \, \tilde{E} \, r_{p}^{2}} \, u^{a}(r_{p}) \, \delta\left(r-r_{p}(t)\right) \, B(l,m) \, e^{-im\phi_{p}(t)}
\label{p of source for point particle} ~,\ee
 where $B(l,m)$ is defined by \be
\left.X^{*lm}_\phi\right|_{\theta=\frac{\pi}{2}}=B(l,m) \, e^{-im\phi}
\label{b def} ~,\ee
 and can be explicitly written as \be
 B(l,m):=-\sqrt{\frac{2l+1}{4\pi}\frac{(l-m)!}{(l+m)!}} \, \frac{(l+m)!!}{(l-m-1)!!}
 \, \sin\left(\frac{\pi}{2}(l+m)\right) \label{b value} ~,\ee for $l>m$
(for $l=m$ replace the factor of $(l-m-1)!!$ in the denominator
with $\sqrt{\pi}$).

Equation (\ref{rw general source term}) gives us an expression for
the odd-parity source term as a function of $r$ and $t$ \be
S_{\rm
o}^{lm}(r,t)=\frac{2r}{(l-1)(l+2)}\left[f^{-1} \, \partial_{t} \, P^{r}+\partial_{r} \, (f \, P^{t})\right]
\label{rw source time domain} ~.\ee

The source term in the frequency domain is given by \be
 S_{\rm o/e}^{lm}(r,\omega)=\int_{-\infty}^{\infty}{e^{-i\omega t} \, S_{\rm o/e}^{lm}(r,t) \, dt}
\label{fourier transform of source term} ~.\ee
 To deal with the time derivative in the integrand of (\ref{fourier transform of source
term}), coming from the first term in (\ref{rw source time
domain}), we integrate by parts to get a factor of $-i\omega$ from
the exponent's time derivative. Now we perform a change of
variables in the integral, $t\to r_{p}(t)$.
Additionally, we have explicit analytical expressions
for the functions $t_p=t_p(r)$ (equation (\ref{t-and-r})),
appearing in the integrand in the factor $e^{-i\omega t_p(r)}$,
and $\phi_p=\phi_p(r)$ (equation (\ref{r and phi})), appearing in
the factor $e^{-im\phi_{p}(t)}$. Using these relations, all the
dependence of the integrand on $r_{p}(t)$ is explicit. Now, using
the delta functions, the integral can be evaluated analytically.

The final result for the frequency domain odd-parity source term
is, for $r_{s}<r<3r_{s}$
 \bea
 S_{\rm o}^{lm}(r,\omega) &=& - \frac{32\sqrt{3} \, \pi \, \tilde{m} \, r_{s}}{\tilde{\Lambda} r} \, B(l,m) \, e^{-i(\omega \, t_p(r)+
 m \, \phi_p(r))}  \\
 &\times& \left[ - i \omega \left( 9 \, \left( \frac{r}{3r_{s} - r} \right)^3 \left( 1+3\left( \frac{r_{s}}{r} \right)^2 \right) \right)- i m \, \frac{6 \sqrt{6} \, r_{s} \, r}{(3r_{s}-r)^{3}}+
 \frac{ \sqrt{2} \, r^{1/2} \, (3r_{s} - 4 r) }{ (3r_{s}-r)^{5/2}} \right]\nonumber
 \label{rw source term - frequency domain}
 ~,\eea
 where $\tilde{\Lambda}=(l-1)l(l+1)(l+2)$.
For $r \notin(r_{s},3r_{s})$, $S_{\rm o}^{lm}(r,\omega) = 0$.

The calculation of $S_{\rm e}^{lm}(r,\omega)$ is done in a very similar
way (but it is a bit longer). Instead of calculating $P^a$ we will need to
calculate the $Q$'s from equations (\ref{q even source
term.1})-(\ref{q even source term.4}). We will also write explicit
expressions for the relevant spherical harmonics in the
$\theta=\frac{\pi}{2}$ plane \bea
 \left.Y^{*lm}\right|_{\theta=\frac{\pi}{2}}&=&A(l,m) \, e^{-im\phi} \\
 \left.Y^{*lm}_\phi\right|_{\theta=\frac{\pi}{2}}&=&-i m \, A(l,m) \, e^{-im\phi} \\
 \left.Y^{*lm}_{\phi\phi}\right|_{\theta=\frac{\pi}{2}}&=&\left(-m^{2}+\frac{1}{2} \, l(l+1)\right)A(l,m) \, e^{-im\phi}
\label{a def} ~,\eea where \be
 A(l,m) := \sqrt{\frac{2l+1}{4\pi}\frac{(l-m)!}{(l+m)!}} \, \frac{(l+m-1)!!}{(l-m)!!} \, \sin \left(\frac{\pi}{2}(l+m+1)\right)
 \label{a value} ~.\ee

After calculating the time domain source term by equation
(\ref{general even source term}), we Fourier transform it, in the
same spirit as in the odd-parity case (using the same change of
variables), to obtain an explicit analytical expression.

The final result is, for $r_{s}<r<3r_{s}$ \bea
S_{\rm e}^{lm}(r,\omega)  &=& -  \frac{16 \pi \, \tilde{m}}{(\mu+2) \Lambda r^{5/2}} \, A(l,m) \, e^{-i(\omega  \, t_p(r)+m \, \phi_p(r))}   \nonumber\\
 &\times&  \left[  \frac{9 r r_{s}^2 \Lambda  \left(-2 m^2+\mu +2\right)}{\mu  (3 r_{s}-r)^{3/2}}+\frac{6 i \sqrt{r} \left(r^2+3 r_{s}^2\right) \left(3 \sqrt{3} m r_{s} (r_{s}-r)- 2 \sqrt{2} r^3 w\right)}{(3
   r_{s}-r)^3}\right. \nonumber\\
   &-& \left.4 i \sqrt{3} m \sqrt{r} r_{s}+\frac{3 \left(r^2+3 r_{s}^2\right) \left(r^2 (\mu -2) \mu +6 r r_{s} (\mu -3)+21 r_{s}^2\right)}{r \Lambda  (3 r_{s}-r)^{3/2}}\right. \\
   &-& \left. \frac{3 (r-r_{s})
   \left(4 r^3-3 r^2 r_{s}+24 r r_{s}^2-45 r_{s}^3\right)}{(3 r_{s}-r)^{5/2}}+\frac{18 r_{s}^2 (r_{s}-r)}{(3 r_{s}-r)^{3/2}}-\frac{4 r_{s} (3 r_{s}-r)^{3/2}}{r \Lambda } \right]\nonumber
  \label{even source term - frequency domain}
 ~.\eea
 For $r\notin(r_{s},3r_{s})$, $S_{\rm e}^{lm}(r,\omega) = 0$.\\
Note that $S_e$ is non-zero only for even $l+m$ and similarly for $S_o$.

%----------------------------------------------------------------------%
%----------------------------------------------------------------------%
 \section{Solving the wave equation}
\label{sec:quasi}
%----------------------------------------------------------------------%
%----------------------------------------------------------------------%

In this section we discuss the solution to the wave equation
(\ref{eqn:wave}).

\subsection{Green's function}
 \label{Integration of the Wave equation}

Knowing the explicit form of the source terms we proceed to
calculate the gravitational waveform. The boundary conditions for equation (\ref{eqn:wave}) are outgoing
waves at infinity \be
 \psi_{\rm o/e}^{lm}(r,\omega) \propto e^{-i \omega r_{*}} ~ ; ~
 r_{*} \to
\infty \label{outgoing at infinity} ~,\ee
 and ingoing boundary
conditions at the horizon \be
 \psi_{\rm o/e}^{lm}(r,\omega)
\propto e^{i \omega r_{*}} ~ ; ~
 r_{*} \to -\infty
 \label{ingoing at horizon} ~.\ee

The solution to equation (\ref{eqn:wave}) is given by \be
 \psi_{\rm o/e}^{lm}(r,\omega)=\int{G_{\rm o/e}^{l}(\omega,r_{*},r'_{*}) \, f \, S_{\rm
o/e}^{lm}(r'_{*},\omega) \, dr'_{*}}
 \label{general solution with green's function} ~,\ee
 where $G_{\rm o/e}^{l}(\omega,r_{*},r'_{*})$
is the frequency domain Green's function given the above-mentioned
boundary conditions. It satisfies \be
 \left[\frac{\partial^2}{\partial r_{*}^{2}}+\omega^{2}-fV_{\rm
o/e}^{l}(r)\right]G_{\rm
o/e}^{l}(\omega,r_{*},r'_{*})=\delta(r_{*}-r'_{*}) \label{green's
function defining equation} ~,\ee
 with the same boundary conditions.

In order to construct the Green's function we must first define two
independent solutions to the homogeneous equation (c.f.
\cite{Arfken}) \be
 \left[\frac{\partial^2}{\partial r_{*}^{2}} + \omega^{2} - f \, V_{\rm o/e}^{l}(r)
 \right]u_{\rm o/e}^{l}(\omega,r_{*})=0
 \label{homogeneous equation} ~.\ee
  Let us denote by $u_{\infty}^{l}$
the solution to (\ref{homogeneous equation}) which satisfies (we
will sometimes suppress the parity) \bea
 u_{\infty}^{l}(r_{*},\omega) \rightarrow e^{ -i\omega  r_\ast}\ ~
; ~ r_{*}\rightarrow  +\infty ~.\ \label{u infinity definition}
\eea At the horizon $u_{\infty}$ can be expanded as follows \be
 u_{\infty}^{l}(r_{*},\omega) \rightarrow A_{\rm out}^{l}(\omega) \, e^{ -i\omega  r_{*}} + A_{\rm
in}^{l}(\omega) \, e^{ i\omega r_{*}} ~ ; ~ r_{*} \rightarrow -\infty ~,
\ \ee
 where $A_{\rm in}$ and $A_{\rm out}$ are some $\omega$
dependent complex coefficients. In a similar manner, we denote by
$u_{\rm hor}^{l}$ the solution to (\ref{homogeneous equation})
which satisfies \be
 u_{\rm hor}^{l}(r_{*},\omega) \rightarrow e^{
i\omega  r_{*}} ~ ; ~ r_{*}\rightarrow  -\infty \
 \label{u horizon definition} ~.\ee
  At infinity it can be expanded as \be
   u_{\rm hor}^{l}(r_{*},\omega) \rightarrow B_{\rm out}^{l}(\omega) \, e^{
-i\omega  r_{*}} + B_{\rm in}^{l}(\omega) \, e^{ i\omega r_{*}} ~ ;
 ~ r_{*} \rightarrow +\infty ~.\  \label{def-Bin} \ee
  The Green's function satisfies
(\ref{green's function defining equation}) with outgoing wave
boundary conditions at infinity. Therefore, for $r_{*}>r'_{*}$ the
solution is proportional to $u_{\infty}^{l}$. Similarly, for
$r_{*}<r'_{*}$ the solution is proportional to $u_{\rm hor}^{l}$.
What remains is to properly match the solutions at $r_{*}=r'_{*}$
and obtain the Green's function
 \be
 G_{\rm o/e}^{l}(\omega,r_{*},r'_{*}) = \left\{
 \begin{array}{ll} W^{-1}(\omega) \, u_{\rm hor}^{l}(r_{*},\omega) \, u_{\infty}^{l}(r'_{*},\omega) & ;
   r_{*}<r'_{*} \\  \\ W^{-1}(\omega) \, u_{\rm hor}^{l}(r'_{*},\omega) \, u_{\infty}^{l}(r_{*},\omega) & ;  r_{*}>r'_{*}  \end{array}
    \right.
 \label{green's function solution}
 ~,\ee
 where $W(\omega):=u_{\rm{hor}}^{l} \partial_{r_{*}} u_{\infty}^{l} - u_{\infty}^{l} \partial_{r_{*}} u_{\rm{hor}}^{l}$
is the Wronskian of $u_{\rm hor}^{l}, u_{\infty}^{l}$ and is
independent of $r_{*}$. Evaluating it both at the horizon and
asymptotically we obtain $W(\omega)=  -2 i\, \omega\, B_{in} =
-2i\, \omega\, A_{out}$ and in particular $B_{in}=A_{out}$.

We are interested in calculating the gravitational waveform
measured by a distant observer ($r \gg r_{s}$). We can, therefore,
simplify the Green's function (\ref{green's function solution})
\be
 G_{\rm o/e}^{l}(\omega,r_{*},r'_{*}) \simeq -\frac{1}{2 i \, \omega \, B_{in}} \, e^{-i \omega r_{*}} \, u_{\rm   hor}^{l}(r'_{*},\omega)
 \label{simplified green's function solution}
~.\ee

Knowing the Green's function (\ref{green's function
solution}), (\ref{simplified green's function solution}) we can
proceed to express the solution to the wave equation. Using
(\ref{general solution with green's function}) and transforming
back to the time domain we have
 \be
 \psi_{\rm o/e}^{lm}(r,t) = \frac{1}{2 \pi} \int_{r_{s}}^{3r_{s}}dr' \left
[\int_{-\infty}^{\infty} e^{i \omega t} \, G_{\rm
o/e}^{l}(\omega,r_{*},r'_{*}) \, S_{\rm o/e}^{lm}(r'_{*},\omega) \, d
\omega\ \right]
 \label{time domain solution for master function} ~.\ee

\subsection{Ringdown amplitudes}
\label{Calculation of QNM Amplitudes}

\begin{figure}
\begin{center}
\includegraphics[width=15cm,height=10cm]{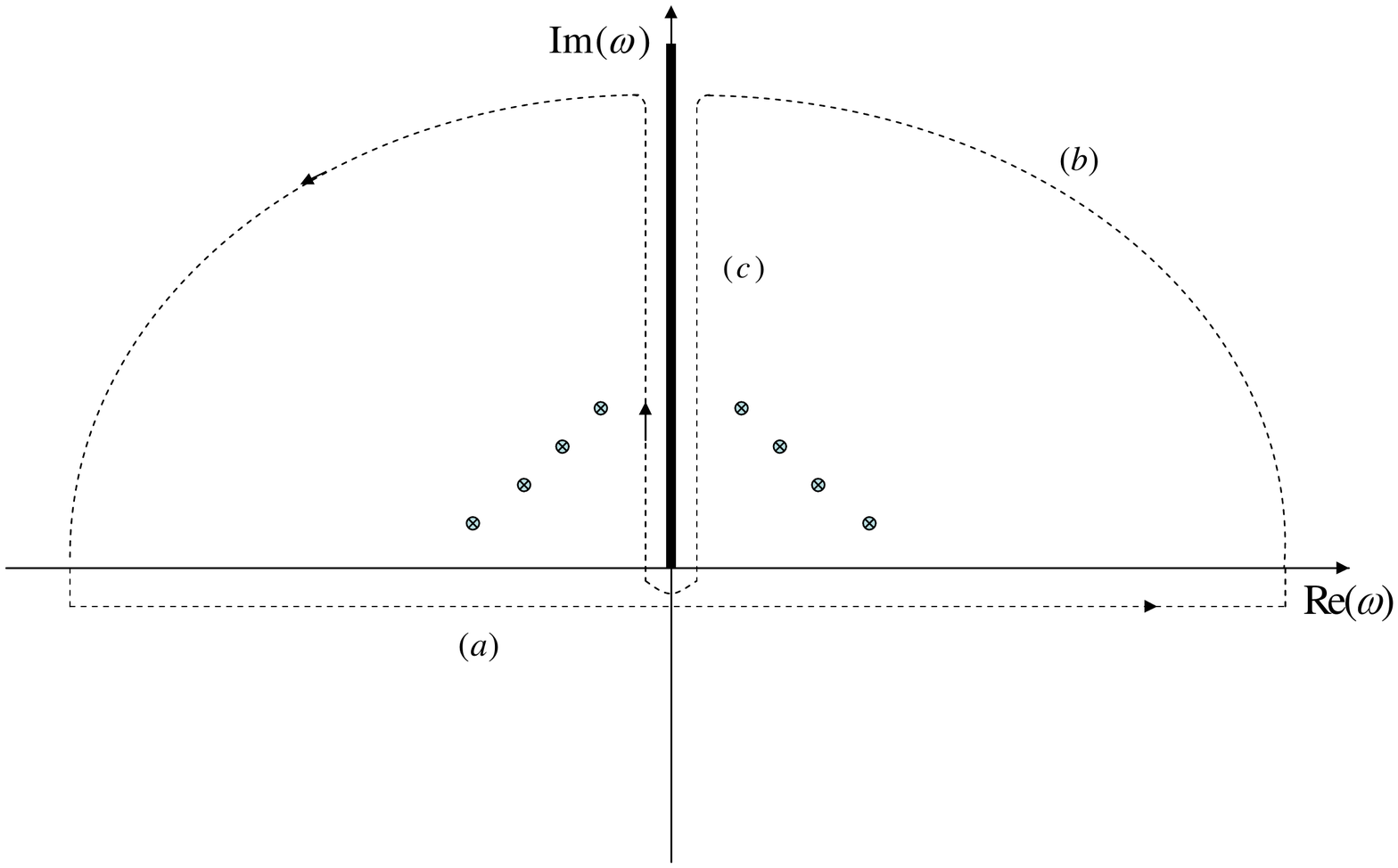}
\caption{Analytic structure of the Green's function in the upper
half $\omega$ plane and the integration contour. The thick line
represents the branch cut on the positive $\rm{Im} (\omega)$ axis.
The dashed line represents the integration contour. The dots
represent some of the Green's function's poles at the QNM
frequencies.} \label{fg:complex plane}
    \end{center}
\end{figure}

It is known that as a perturbed black hole settles back to its
stationary state the background metric exhibits an exponentially
decaying radiation known as ringdown, or quasi-normal modes. This
can be seen from the analytic structure of Green's functions for
the Regge-Wheeler and Zerilli equations in the complex frequency
domain. It is known (e.g.\cite{Leaver:1986gd}) that these Green's
functions have poles in the upper half plane ($\rm{Im}(\omega) >
0$) and a branch cut along the positive part of the
$\rm{Im}(\omega)$ axis as shown schematically in figure
\ref{fg:complex plane}.  The poles are interpreted as the
quasi-normal modes. Their locations are the quasi-normal
frequencies, whose imaginary part is responsible for their
decaying nature at late times. The branch cut is interpreted as
the ``tail'' resulting from scattering the waves off the
background geometry at large radii.
Here we shall proceed to derive a formula for the ringdown or
quasi-normal mode amplitudes. Concentrating on them is justified
since they are known to dominate over the tail for late times but
not too late times -- time region (iv) of \cite{Leaver:1986gd},
the discussion after eq. (46).
%Far away the tail's amplitude is suppressed by powers of $r_s/r$
% relative to the leading ringdown signal and therefore will be
% ignored from hereon.
See \cite{Andersson:1996cm,Ching:1995tj,HodPiran,BarackOri} for
further discussion.
In addition, these amplitudes are one of the ingredients which go
into current waveform models, as reviewed in the introduction, but
as of now they are matched (with a different part of the modeled
waveform) rather than computed from the theory.

%----------------------------------------------------------------------%
%----------------------------------------------------------------------%

We now turn to evaluate the integral in square brackets in
(\ref{time domain solution for master function}), utilizing the
continuation to complex frequencies. We deform the contour from
the real axis to parts (b) and (c) in figure \ref{fg:complex
plane} and collect the residue contribution from the poles.

The poles correspond to zeros of the Wronskian. We denote by
$\omega_{n l}$ the $n$-th pole for a given $l$ (ordered by
increasing imaginary part).  At these frequencies $B_{\rm
in}(\omega_{n l})= A_{\rm out}(\omega_{n l})=0$ and the leading
behavior for $B_{\rm in}$ is \be
 B_{\rm in} \simeq  \left. \partial_{\omega} B_{\rm in} \right|_{\omega = \omega_{nl}} ~ (\omega - \omega_{nl})
\label{definition of beta}
 ~.\ee
For convenience we assign notation to these constants \be
 \beta_{nl} := \left. -\partial_{\omega} B_{\rm in} \right|_{\omega =
\omega_{nl}}
 \label{definition of beta1} ~.\ee
Their formalism and numerical computation can be found at
\cite{BertiCardoso06}.

The $r_{*} \gg r_{s}$ Green's function (\ref{simplified green's
function solution}) (implying $r_{*}> r'_{*}$) for $\omega \approx
\omega_{nl}$ can now be written as \be
 G_{\rm o/e}^{l}(\omega,r_{*},r'_{*}) = \frac{e^{-i \omega_{nl}
r_{*}} u_{\rm hor}^{l}(r'_{*})}{2 i \, \omega_{nl} \, \beta_{nl}
(\omega - \omega_{nl})}
 \label{green's function at qnm at large r star} ~.\ee

At late times when the contribution of parts (b) vanishes and part
(c) is neglected as we discussed above the integral over
frequencies (\ref{time domain solution for master function}) can
be approximated by the sum over the residues \be
 \psi_{\rm o/e}^{lm}(r,t) = \frac{1}{2 \pi} \int_{r_{s}}^{3r_{s}}dr' \sum_{n}
2 \pi i \left. \rm{Res} \right|_{\omega = \omega_{nl}} \left [
e^{i \omega t} \, G_{\rm o/e}^{l}(\omega,r_{*},r'_{*}) \, S_{\rm
o/e}^{lm}(r'_{*},\omega) \right]
 \label{time domain solution for
master function-using cauchy's theorem} ~.\ee
 Plugging in (\ref{green's function at qnm at large r star}), we obtain \be
\psi_{\rm o/e}^{lm}(r,t)= \sum_{n} \frac{ e^{i \omega_{nl}
(t-r_{*})}}{2 \, \omega_{nl} \, \beta_{nl}}\int_{r_{s}}^{3r_{s}}u_{\rm
hor}^{l}(r'_{*}) \, S_{\rm o/e}^{lm}(r'_{*},\omega_{nl}) \, dr'
 \label{time domain solution for master function-final expression} ~,\ee
  where the index $n$ runs over all QNM's for a given $l$.

We identify the coefficient of $e^{i \omega_{nl} (t-r_{*})}$ with
the ringdown amplitude \be
 R_{nlm} := \frac{1}{2 \, \omega_{nl} \, \beta_{nl}}\int_{r_{s}}^{3r_{s}}u_{\rm hor}^{l}(r'_{*}) \, S_{\rm
o/e}^{lm}(r'_{*},\omega_{nl}) \, dr'
 \label{amplitude def} ~.\ee
This is our main result for the ringdown amplitudes.

For $r \to \risco$ $u_{\rm hor}$ is finite and the third order poles in the source $S$ are suppressed by the term $\exp(-i \omega t)$ which is bounded by
$|\exp(-i \omega t)|  \le \exp ({\rm Im}(\omega)t) \simeq \exp [{\rm Im}(\omega)\cdot const /\sqrt{r-\risco}]$ and so the integrand vanishes exponentially
in the limit.

In the limit $r \to r_{\rm hor}$, however, the wavefunction diverges
\be
 u_{\rm hor} \propto \exp(i \omega r_{*}) \propto \(\frac{r}{r_{s}}-1\)^{i
\frac{\omega}{r_{s}}}.
\ee
The source term, too, diverges in this limit in the same fashion. At $r \to r_s$ the trajectory satisfies $r_* \simeq t + const$
and so
\be
S_{o/e} \propto \exp(-i \omega t) \propto \exp(i \omega r_{*}) \propto \(\frac{r}{r_{s}}-1\)^{i \frac{\omega}{r_{s}}}.
\ee
Altogether, then, the
integrand is $\propto \(\frac{r}{r_{s}}-1\)^{2 i \frac{\omega}{r_{s}}}$ and so the integral diverges when ${\rm Im}(\frac{\omega}{r_{s}}) \geq
\frac{1}{2}$. This spurious divergence needs to be regularized. This problem of a diverging QNM excitation integral was realized already in
\cite{Leaver:1986gd} for different excitation scenarios. Two different methods (equivalent in principle) to regularize the integral are discussed there, one of them
being the divergence subtraction method (introduced in \cite{Detweiler:1979xr} for excitation integrals of non-QNM frequencies). This is the method we used
in practice when evaluating the amplitudes numerically, and it will be discussed in the next section.

%----------------------------------------------------------------------% %----------------------------------------------------------------------%
 \section{Numerical Evaluation}
\label{sec:numerics}
%----------------------------------------------------------------------% %----------------------------------------------------------------------%
In this section we numerically evaluate the amplitudes (\ref{amplitude def}).

\subsection{Method}
 \label{numerical method}

We started by calculating the numerical values of the QNM
frequencies $\omega_{nl}$. One can use either one of several
methods (see the review \cite{QNM-rev}), and we chose Leaver's
continued fraction method (see \cite{Leaver:1986gd}) which can be
extended to yield also the QNM wavefunctions $u_{\rm
hor}(\omega_{nl},r)$. More specifically, the wavefunctions are
obtained in the regime of interest as a power-series expansion in
$(1-\frac{r_{s}}{r})$ multiplying a factored-out non-analytical
part. The calculated frequencies were confirmed by the literature,
and the wavefunctions were confirmed by  substitution into the
homogeneous Regge-Wheeler and Zerilli equations. We continued to
calculate the excitation factors (or transmission residues)
$\beta_{nl}$, again using Leaver's method, and the first few $n$'s
and $l$'s were confirmed by Leaver's results. We found
\cite{BertiCardoso06} to be a very useful and detailed guide to
the calculation of the excitation factors (there the more general
rotating BH case is considered).

The next ingredient in the integrand is the source term $S_{\rm
o/e}^{lm}(r_{*},\omega_{nl})$, which is given analytically by (\ref{rw source term - frequency domain}), (\ref{even source term - frequency domain}) where one should substitute
the trajectory functions $t_p(r)$ (\ref{t-and-r}) and $\phi_p(r)$ (\ref{r and phi}).

It remains to perform the numerical integration in (\ref{amplitude
def}). As noted in the end of last section, this integral diverges
and must be regularized by what is essentially an analytic
continuation. To do so, we used the so-called divergence
subtraction method (\cite{Detweiler:1979xr},\cite{Leaver:1986gd}).
The idea is the following: we set the integrand to zero at the
horizon ($r \to r_{\rm hor}$) by subtracting from it a function
$f$ that behaves the same as the integrand as $r \to r_{\rm hor}$,
thereby getting rid of the unphysical divergence at that
integration limit. In order not to add anything finite to the
integral we will need to add the value of $F := \int{f dr}$ at the
upper limit of integration. The integration constant in $F$ is
determined by the condition that $F$ does not have a part $\sim
const.$ as $r \rightarrow 1$. That is, if $f$ is of the form given
below in (\ref{f definition}), we integrate it using the rule
$\int r^{\alpha} dr = \frac{r^{\alpha + 1}}{\alpha + 1}$ without
the addition of a constant. In summary \be
 \left. \int_{r_{s}}^{3r_{s}}u_{\rm hor}^{l}(r'_{*}) \, S_{\rm
o/e}^{lm}(r'_{*},\omega_{nl}) \, dr' \right|_{physical \ part} = \int_{r_{s}}^{3r_{s}} \[ u_{\rm hor}^{l}(r'_{*}) \, S_{\rm o/e}^{lm}(r'_{*},\omega_{nl}) -f
\] \, dr' + \left. F \right|_{3r_{s}}
 \label{regularization of integral} ~.\ee

$f$ is, of course, not unique. Any function which will be easy to integrate analytically (and behaves in the desirable way as $r \to r_{\rm hor}$) will do.
The (natural) $f$ we chose to use was
\be
f=\(\frac{r}{r_{s}}-1\)^{2 i \frac{\omega}{r_{s}}} \(A_{0}+(A_{1} (r-r_{s})+...\)
\label{f definition}
\ee
where the coefficients $A_{i}$
are determined from our expressions for $u_{\rm hor}^{l}$ and $S_{\rm o/e}^{lm}$ and enough terms must be taken as to assure convergence.

%----------------------------------------------------------------------%
%\begin{figure}[!h]
%\begin{center}
%\includegraphics[width=15cm,height=8cm]{ringdown_waveform}
%\caption{The ringdown waveform, as calculated from the ringdown amplitudes. Here it is assumed the observer is situated in the equatorial plane, at $\phi=0$, %$\theta=\frac{\pi}{2}$. The horizontal axis represents $t-r$ in units of $r_{s}$, while the vertical axis represents the magnitude of $h_{+}$, the $+$ polarized %part of the gravitational wave.} \label{fg:ringdown waveform}
%    \end{center}
%\end{figure}
%----------------------------------------------------------------------%

%----------------------------------------------------------------------%
\begin{center}
\begin{figure*}[b!]
%\setlength{\tabcolsep}{ 40 pt }
%\footnotesize{
        \begin{tabular}{cc}
            \includegraphics[width=16cm,height=7cm]{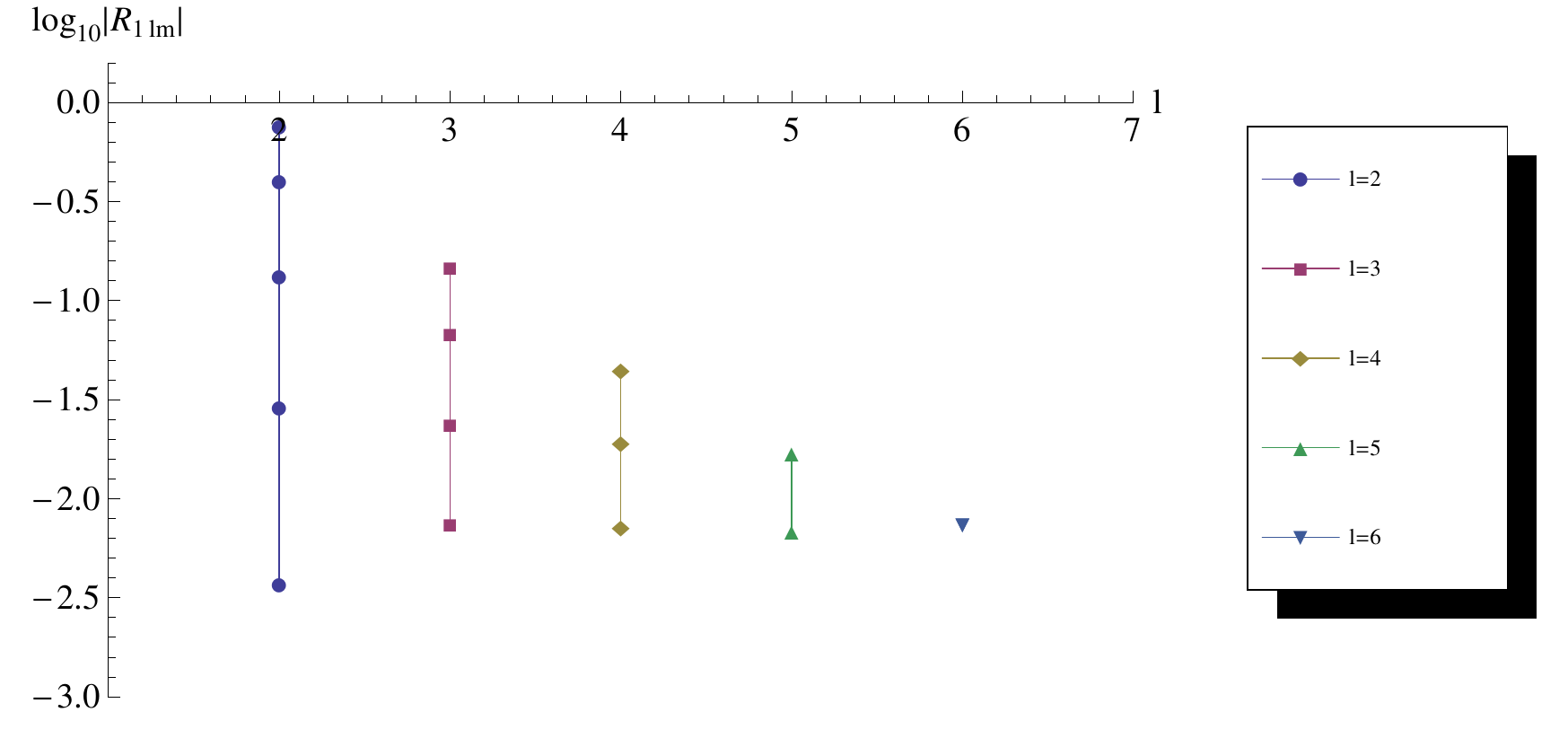} &
        \end{tabular}
%}
 \caption{Ringdown amplitudes from tables 1,2 depicted on a logarithmic
 scale for leading $n=1$, $2 \le l \le 6$ and several $m$ values. The amplitudes decrease as $l-m$ increases, for instance for $l=3$ the
  top point corresponds to $m=3$ while the bottom one to $m=0$.}
 %The horizontal axis is $l$ while the vertical axis is  $\log_{10}|R|$.}
 \label{fg:amplitude listplot}
\end{figure*}
\end{center}
%%----------------------------------------------------------------------%

%----------------------------------------------------------------------%
\begin{center}
\begin{figure*}[b!]
%\setlength{\tabcolsep}{ 40 pt }
%\footnotesize{
        \begin{tabular}{cc}
            \includegraphics[width=15cm,height=7cm]{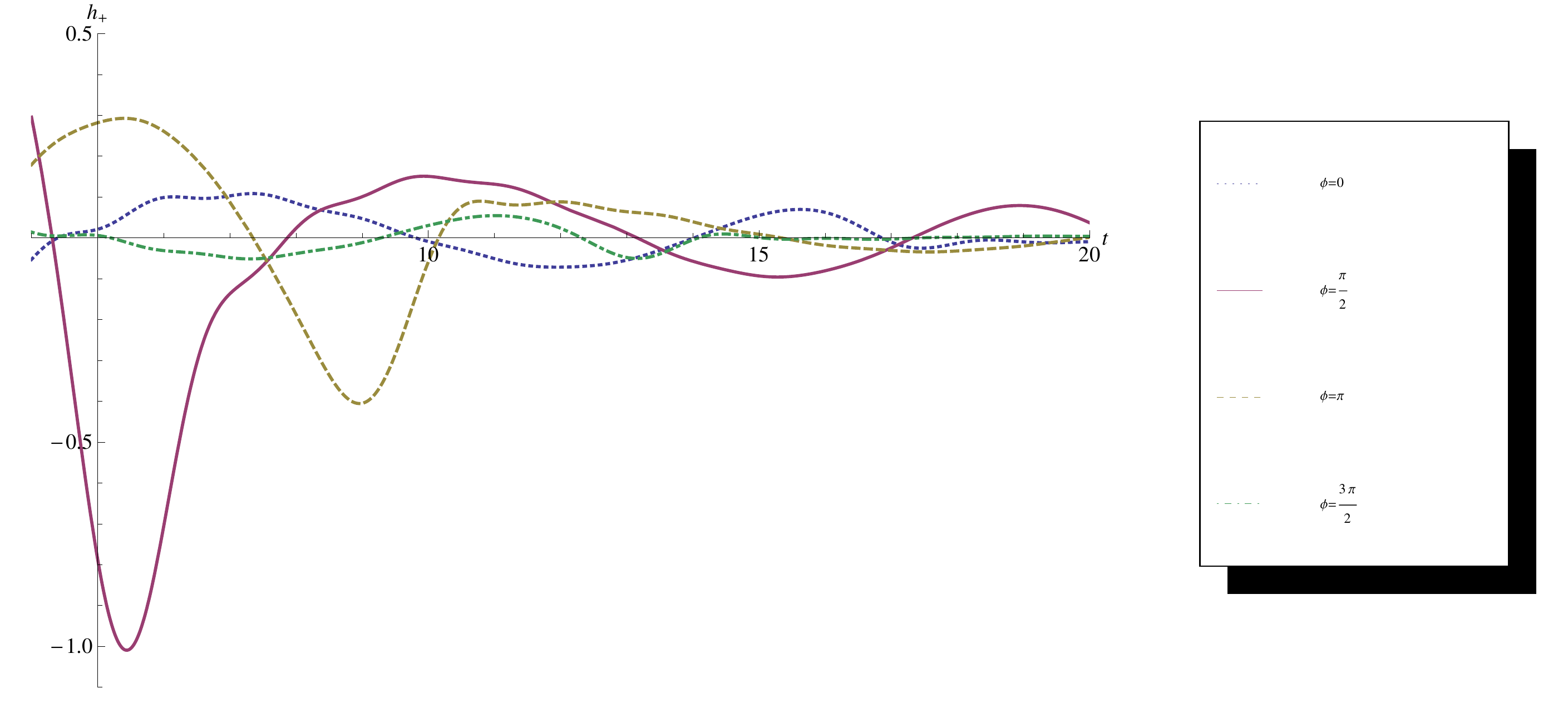} &
        \end{tabular}
%}
 \caption{Ringdown waveforms summing the dominant $(l,m,n)$ contributions.
 The (distant) observer is situated in the equatorial plane ($\theta=\frac{\pi}{2}$), at: (a) $\phi=0$, (b) $\phi=\frac{\pi}{2}$, (c) $\phi=\pi$, (d) $\phi=\frac{3 \pi}{2}$. The horizontal axis represents $t-r$ in units of $r_{s}$, while the vertical axis represents $h_{+}$, the $+$ polarized part of the gravitational wave normalized by $r/\tilde{m}$. Notice the asymmetric nature of the ringdown waveform.} \label{fg:ringdown waveform}
\end{figure*}
\end{center}
%%----------------------------------------------------------------------%

\subsection{Results}
 \label{numerical results}

From equation (\ref{time domain solution for master function-final expression}) it is clear that the ringdown amplitudes are defined only up to a shift in the time coordinate; if we take $t \to t + t'$ the amplitudes will transform as
\bea
R_{nlm} \to R_{nlm} e^{i \omega_{nl} t'}.
\label{amp transformation}
\eea

We fix $t_0$, the integration constant in equation
(\ref{t-and-r}), somewhat arbitrarily such that $r(t=0)=1.1\,
r_s$, which we observed to be shortly after the onset of ringdown
\cite{InProgress}, namely the region which is well described by
(\ref{time domain solution for master function-using cauchy's
theorem}).

It is worthwhile to note that the decay constants (imaginary parts
of the frequencies $\omega_{nl}$) do not vary with $m$ and vary
only little with $l$ while they vary strongly with overtone number
$n$  and hence (\ref{amp transformation}) shifting $t_0$ will not
affect the ratio of magnitudes of amplitudes which differ only by
their $m$, it will weakly affect different $l$'s and finally
strongly affect the ratio for different $n$'s. Note also that the
real signal is gotten by adding the complex conjugate to sum of
ringdown exponentials. This is equivalent to the doubling of QNM
frequencies $\Re (\omega) \to - \Re (\omega)$ as in figure
\ref{fg:complex plane}.

Results for amplitudes of the first few dominant modes are
displayed in table (\ref{table of amps l=2-5}). In table
(\ref{table of amps l=6-15}) the contribution of higher $l$'s is
displayed, up to $l=10$. In table (\ref{table of frequencies})
numerical values of QNM frequencies are displayed for the first
few $(n,l)$. Figure \ref{fg:amplitude listplot} gives a graphic
illustration of several amplitudes on a logarithmic scale.

In figure \ref{fg:ringdown waveform} we display the ringdown
waveform (calculated from the amplitudes through equations
(\ref{time domain solution for master function-using cauchy's
theorem}), (\ref{metric perturbation at null infinity - AB})), as
viewed by observers in different angular positions. Notice the
highly asymmetric nature of the ringdown waveform - observers in
different azimuthal positions see very different waveforms. This
is due to the fact that the trajectory breaks azimuthal symmetry,
by infalling at some specific angular position. Of course,
waveforms for observers in any angular position can be easily
calculated from the amplitudes, via equation (\ref{metric
perturbation at null infinity - AB}).

\begin{table}[!h]
\centering \caption{ \label{table of amplitudes 1} Numerical values of the ringdown amplitudes $R_{nlm}$ for $l=2,3,4,5$ , several values of the  overtone number $n$ and all values of $m$. Conventions: $t_0$ (2.10) %(\ref{t-and-r})
was taken such that the plunging object is at $1.1\, r_s$ at $t=0$, the overall  normalization is defined by (3.10,3.13),
and the normalization of spherical harmonics is given in footnote 7.} %(\ref{cpm function},\ref{zm function}).}
\begin{tabular}{cccccc}
%\begin{tabular}{||c|c|c|c|c|c|c||}  \hline
\multicolumn{4}{c}{$l=2$} \\
\hline
 $m$ &$n=1$   &$n=2$    &$n=3$     \\
\hline
$2$ & $-0.0985724 - 0.747787 i$& $-0.229354 + 0.428849 i$& $0.167484 - 0.251937 i$ \\
$1$ & $-0.0210521 + 0.399297 i$& $0.441304 - 0.31877 i$& $-0.489451 + 0.053276 i$ \\
$0$ & $-0.0887841 + 0.0979244 i$ & $0.303357 + 0.0416042 i$& $-0.249012 - 0.314569 i$ \\
$-1$ & $0.0274099 + 0.00889306 i$& $-0.0324417 - 0.0903849 i $ & $-0.0867015 + 0.154488 i$ \\
$-2$ & $(-0.735088 + 3.59504 i)\times 10^{-3}$ & $0.0135078 - 0.0089118 i$ & $-0.0394189 - 0.010878 i $\\
\hline
\hline
\multicolumn{4}{c}{} \\
\multicolumn{4}{c}{$l=3$} \\
\hline
 $m$ &$n=1$   &$n=2$    &$n=3$     \\
\hline
$3$ & $-0.0375476 + 0.141024 i$& $0.117107 - 0.057957 i$& $-0.0862334 + 0.00839625 i$ \\
$2$ & $0.0336116 - 0.0585892 i$& $-0.104903 + 0.00738472 i$& $0.0855561 + 0.0540601 i$ \\
$1$ & $0.0186212 - 0.0145793 i$ & $-0.0499562 - 0.0209467 i$& $0.0245107 + 0.0656791 i$ \\
$0$ & $(-7.36412 + 0.205584 i)\times 10^{-3}$ & $0.0119392 + 0.0185584 i$ & $0.0108096 - 0.0349259 i$ \\
$-1$ & $(-1.32958 - 1.74202 i)\times 10^{-3}$ & $-2.4849 + 7.94279 i)\times 10^{-3}$ & $0.0160937 - 0.00679286 i$\\
$-2$ & $(-3.59688 + 3.99339 i)\times 10^{-4}$ & $(2.43071 + 0.097078 i)\times 10^{-3}$ & $(-3.98823 - 4.75782 i)\times 10^{-3}$\\
$-3$ & $(-5.2293 - 6.09408 i)\times 10^{-5}$ & $(-0.284844 + 4.19003 i)\times 10^{-4}$ & $(9.4908 - 8.54244 i)\times 10^{-4}$\\
\hline
\hline
\multicolumn{4}{c}{} \\
\multicolumn{4}{c}{$l=4$} \\
\hline
 $m$ &$n=1$   &$n=2$      \\
\hline
$4$ & $0.0318095 - 0.0308272 i$& $-0.0489266 - 0.00923324 i$ \\
$3$ & $-0.0165716 + 0.00919402 i$& $0.0289206 + 0.0167393 i$ \\
$2$ & $(-6.86211 + 1.83095 i)\times 10^{-3}$ & $0.0114492 + 0.0128208 i$ \\
$1$ & $(2.32275 + 0.593617 i)\times 10^{-3}$ & $(-1.76906 - 7.17998 i)\times 10^{-3}$  \\
$0$ & $(4.47817 + 5.89624 i)\times 10^{-4}$ & $(1.10773 - 2.61122 i)\times 10^{-3}$ \\
$-1$ & $(0.514994 - 2.12774 i)\times 10^{-4}$ & $(-9.22639 + 3.19073 i)\times 10^{-4}$ \\
$-2$ & $(6.24213 - 1.42382 i)\times 10^{-5}$ & $(-2.49301 - 2.12194 i)\times 10^{-4}$ \\
$-3$ & $(-0.482705 - 1.61408 i)\times 10^{-5}$ & $(-5.20619 + 8.05728 i)\times 10^{-5}$ \\
$-4$ & $(2.7626 - 0.0853029 i)\times 10^{-6}$ & $(-1.26026 - 1.17755 i)\times 10^{-5}$ \\
\hline
\hline
\multicolumn{4}{c}{} \\
\multicolumn{4}{c}{$l=5$} \\
\hline
 $m$ &$n=1$        \\
\hline
$5$ & $-0.0166349 + 0.00302628 i$ \\
$4$ & $(6.80123 + 0.509818 i)\times 10^{-3}$ \\
$3$ & $(2.61522 + 0.629434 i)\times 10^{-3}$  \\
$2$ & $(-7.32533 - 5.82304 i)\times 10^{-4}$   \\
$1$ & $(-1.14018 - 2.79764 i)\times 10^{-4}$  \\
$0$ & $(-3.06403 + 8.96566 i)\times 10^{-5}$  \\
\hline
\hline
\multicolumn{4}{c}{} \\
\end{tabular}
\label{table of amps l=2-5}
\end{table}

\begin{table}[!h]
  \centering \caption{\label{table of amplitudes 2} Numerical values of the ringdown amplitudes $R_{nlm}$ for $l=6$ to
 $10$, overtone number $n=1$ and dominant $m$'s.}
\begin{tabular}{cccccc}
%\begin{tabular}{||c|c|c|c|c|c|c||}  \hline
\multicolumn{3}{c}{} \\
\hline
$l$  &  $m=l$   &$m=l-1$     \\
\hline
6 & $(6.73817 + 3.1156 i)\times 10^{-3}$& $(-2.21976 - 1.78938 i)\times 10^{-3}$ \\
7 & $(-1.75245 - 3.12505 i)\times 10^{-3}$& $(0.356901 + 1.26881 i)\times 10^{-3}$\\
8 & $(-0.229465 + 1.83699 i)\times 10^{-3}$& $(2.27002 - 6.14468 i)\times 10^{-4}$\\
9 & $(6.97897 - 7.26025 i)\times 10^{-4}$& $(-2.87291 + 1.88928 i)\times 10^{-4}$\\
10 & $(-4.03242 - 3.58438 i)\times 10^{-4}$& $(1.04411 + 1.44422 i)\times 10^{-4}$\\
 \\
\hline
\hline
\multicolumn{3}{c}{} \\
\end{tabular}
\label{table of amps l=6-15}
\end{table}

\begin{table}[!h]
  \centering \caption{Numerical values of some dominant QNM frequencies in units of $r_s^{-1}$.}
\begin{tabular}{cccccc}
%\begin{tabular}{||c|c|c|c|c|c|c||}  \hline
\multicolumn{4}{c}{} \\
\hline
 $l$ &$n=1$   &$n=2$    &$n=3$     \\
\hline
2 & $-0.74734 + 0.17792 i$ & $-0.69342 + 0.54783 i$ & $-0.60210 + 0.95655 i$ \\
3 & $-1.19889 + 0.18540 i$ & $-1.16529 + 0.56259 i$ & $-1.10337 + 0.95818 i$\\
4 & $-1.61836 + 0.18832 i$ & $-1.59326 + 0.56866 i$ & $-1.54542 + 0.95981 i$\\
5 & $-2.02459 + 0.18974 i$ & $$ & $$\\
6 & $-2.42402 + 0.19053 i$ & $$ & $$\\
 \\
\hline
\hline
\multicolumn{4}{c}{} \\
\end{tabular}
\label{table of frequencies}
\end{table}

\subsection*{Acknowledgements}

We thank Amos Ori, Umpei Miyamoto, Omer Bromberg and Gerhard Sch\"{a}fer for
comments on the manuscript.
 We thank Emanuele Berti and Vitor Cardoso for collaboration on a related paper \cite{InProgress}.

This research is supported by The Israel Science Foundation grant
no 607/05, by the German Israel Cooperation Project grant DIP
H.52, and the Einstein Center at the Hebrew University.

\end{document}